\documentclass{moriond}

\usepackage{pifont}
\usepackage{pstricks,epsfig,amssymb}
\def\be{\begin{equation}}
\def\ee{\end{equation}}
\def\bea{\begin{eqnarray}}
\def\eea{\end{eqnarray}}

\newcommand{\cmark}{\ding{51}}

\newcommand{\Photo}{}

\begin{document}
\vspace*{4cm}
\title{LHC HIGGS CP SENSITIVE OBSERVABLES IN $H \to \tau^+\tau^-; \tau^\pm \to (3\pi)^\pm \nu$ \\ AND MACHINE LEARNING BENEFITS}

\author{ E. Richter-Was$^1$ and Z. Was$^2$ }

\address{$^1$ Institute of Physics, Jagiellonian University,  30-348  Krak\'ow, Lojasiewicza 11, Poland \\
  $^2${\it Speaker}, Institute of Nuclear Physics, Polish Academy of Sciences, 31-342 Krak\'ow, Radzikowskiego 152, Poland
  }

\maketitle

\abstracts{ In phenomenological preparation  for new
  measurements one searches for the carriers
  of quality signatures.
  Often, the first approach quantities may be
  difficult to measure or to provide sufficiently precise  predictions
  for comparisons.
  Complexity  of
  necessary details grow with precision.
  To achieve the goal one can not break the theory principles, and take
  into account effects which could be ignored earlier.
  %could have been compromised when precision requirements were less strict.
  Mixed approach where dominant effects are taken into account with
  intuitive even simplistic approach was developed.  Non dominant
  corrections were controlled with the help of Monte Carlo simulations.
  Concept of {\it Optimal
  variables} was successfully applied for many measurements.\\
  New techniques, like Machine Learning, offer solutions to exploit
  multidimensional signatures. Complementarity of these new and old approaches
 is studied for the example of
  Higgs Boson CP-parity  measurements in
  $H \to \tau^+\tau^-$, $\tau^\pm \to \nu (3\pi)^\pm$ cascade decays.\\
  \centerline{IFJPAN-IV-2019-6}\\
  \centerline{May 2019}\\
  \centerline{Presented at Rencontres de Moriond, {\it QCD and High Energy Interactions session}, March 2019}
}

\section{Introduction}

Despite multidimensional nature of high energy measurements, where
big samples of events consisting of observed particles sets are analyzed, it was
generally believed\cite{Davier:1992nw}
that the best for phenomenology purposes is to construct single, one dimensional distribution which is sensitive
to particular quantity of physics interest, such as 
coupling constants, masses or widths of the investigated particles.
Very successful high precision  LEP measurements\cite{ALEPH:2005ab}  were following this approach.

Also for Higgs boson CP parity measurement
such one dimensional {\it Optimal Variable} could have been constructed for
$H\to \tau^+\tau^-$, $\tau^\pm \rho^\pm \nu$, $\rho^\pm \pi^\pm \pi^0$ cascade
decay\cite{Bower:2002zx}. Simulations necessary
to evaluate experimental conditions were performed with the help of
$\tau$ decay Monte Carlo program {\tt TAUOLA}\cite{Jadach:1993hs} and its {\tt universal interface}\cite{Davidson:2010rw}.
Such measurements are feasible, but suffer because small branching
fraction, the  $\tau^\pm \rho^\pm \nu$ contributes only 6.5\%
of $H\to \tau^+\tau^-$ final states. 

For this observable,there was no need to rely on reconstruction
of difficult to constraint with the measurements neutrino momenta.
Each $\tau$ lepton decay channel has different decay products and
distinct detector response.
In\cite{Kuhn:1995nn} it was pointed, that
every $\tau$ decay channel has  the same $\tau$ spin sensitivity.
This requires non-detectable neutrinos to be resonstucted.
What are the possible ways out?
Steps in that directions were attempted already long time
ago\cite{Rouge:2005iy,Desch:2003mw}, but were succesfull only in part.

\subsection{Basic formulation}
Let us explain very briefly the physics context of the problem. 
Higgs  boson Yukawa coupling expressed with the help of the scalar--pseudoscalar mixing
angle $\phi$ reads as
\begin{equation}
  {\cal L}_Y= N\;\bar{\tau}h(\cos\phi+i\sin\phi\gamma_{5})\tau
\end{equation}
where
$N$ denotes normalization and $\bar\tau$, $\tau$ spinors of the $\tau^+$ and $\tau^-$.
The decay probability 
of the scalar/pseudoscalar Higgs
\begin{equation}
  \Gamma(H/A\to \tau^{+}\tau^{-}) 
\sim 1-s^{\tau^{+}}_{\parallel}  
s^{\tau^{-}}_{\parallel}\pm s^{\tau^{+}}_{\perp}s^{\tau^{-}}_{\perp},
\end{equation}
is sensitive to the $\tau^\pm$ polarization   
vectors $s^{\tau^{\pm}}$  (defined    
in their rest frames).    
The symbols {\scriptsize ${\parallel}$,${\perp}$} denote components
 parallel/transverse   
to the Higgs boson momentum as seen from the respective frames.
When   decay 
into $\tau^+ \tau^-$ pair is taken into account, polarization vectors
$s^{\tau^{-}}$ are replaced with
polarimetric vectors $h_\pm$ representing $\tau^\pm$ decay matrix elements.
The $R$ matrix  depicts spin state of the $\tau$ lepton pair.
Formula for the most general mixed parity 
of  $H \to \tau^+\tau^-$ and $\tau^\pm$ decays
can be thus expressed as
\begin{equation} \label{eq:matrix}
  |M|^2\sim 1 +  h_{+}^{i} \cdot  h_{-}^{j} R_{i,j}; \;\;\;\;\; i,j=\{x,y,z\}.
\end{equation}

In notation of
Ref.\cite{Desch:2003rw},
 the corresponding CP sensitive spin weight $wt$ is rather simple:
\begin{equation} \label{eq:wt}
wt = 1-h_{{+}}^{z} h_{{-}}^{z}+ h_{{+}}^{\perp} R(2\phi)~h_{{-}}^{\perp}.
\end{equation}
The formula is valid for $h_\pm$ defined in $\tau^\pm$ rest-frames. The
$R(2\phi)$ denote  the $2\phi$ angle rotation matrix  around the $z$ direction:
$R_{xx}= R_{yy}={\cos2\phi}$, $R_{xy}=-R_{yx}={\sin2\phi}$.
The $\tau^\pm$ decay polarimetric vectors $h_{+}^i$,  $h_{-}^j$, in the simplest case
of $\tau^{\pm} \to \pi^{\pm} \pi^0 \nu $ decay read 
\begin{equation}
h^i_\pm =  {\cal N} \Bigl( 2(q\cdot p_{\nu})  q^i -q^2  p_{\nu}^i \Bigr), \;\;\;  
\end{equation}
where 4-momenta of $\tau$ decay products  $\pi^{\pm}$, $\pi^0$ and $\nu_{\tau}$
are denoted respectively as
$ p_{\pi^{\pm}}$, $p_{\pi^0}, p_{\nu}$ :
%q--> p_{\pi^\pm} -p_{\pi^0}.
The $q=p_{\pi^\pm} -p_{\pi^0}$. Obviously, complete CP sensitivity can be extracted
only if $p_{\nu}$ is controlled.

\subsection{The $\tau$ decay channel independent features.}

Note that  spin weight (\ref{eq:wt}), is a simple 
trigonometric polynomial in  Higgs CP parity mixing angle
$\phi$.
This observation  is valid for all $\tau$ decay channels and that opens
possibility for  studies, where all effort on experimental reconstruction
is concentrated on measurement of 
the polarimetric vectors $h_i$. Final analysis of observable significance
rely on 
(\ref{eq:wt}). Such a path was already followed,
for CMS experiment. 
Preliminary  effort was presented already
at 2018 $\tau$-lepton conference\cite{Cherepanov:2018yqb}. The general principle is of course
much older, see eg.\cite{Rouge:2005iy} and was revisited in\cite{Berge:2015nua}.

\subsection{Multi dimensional nature of the signatures.}

\begin{figure}
  
\begin{minipage}[l]{0.63\linewidth}
%  \centerline{\includegraphics[width=7.5cm,angle=0]{observabla}}
  \epsfig{file=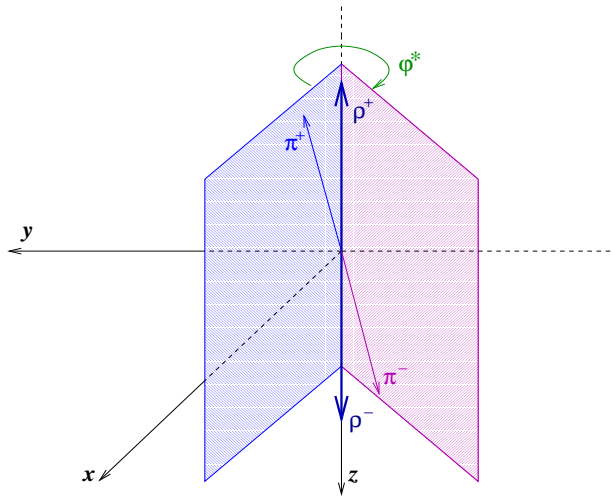,width=80mm,height=65mm}
\end{minipage}
\begin{minipage}[r]{0.33\linewidth}
  \caption{Definition of acoplanarity angle $\varphi^*$ in $\rho^+\rho^-$ pair
    rest-frame by galf-planes spanned over momenta of each $\rho$ visible decay
    products. 
    In case of $\tau^\pm \to (3\pi)^\pm \nu $ four such planes can be defined
    and thus
  number of possible acoplanarity angles increases to four (or sixteen
  if both $\tau^\pm $ decay to $3\pi $).}
\label{fig:acopl}
\end{minipage}
\end{figure}
In\cite{Bower:2002zx} to control parity of the Higgs boson use of the acoplanarity angle
was proposed. Such a  definition for
$\tau^\pm \to \nu \pi^\pm \pi^0$ rely on directly observable
four-momenta, see Fig. \ref{fig:acopl}, only.
The distributions were clearly distinct for scalar
and mixed parity Higgs, see Fig. \ref{fig:acodistr}. To achieve sensitivity the events had to be separated
to two groups accordingly to the sign of the product $y_1 y_2$, where
\begin{equation} \label{eq:yi}
y_1={E_{\pi^{+}}-E_{\pi^{0}}\over E_{\pi^{+}}+E_{\pi^{0}}},\;\;\;\;
y_2={E_{\pi^{-}}-E_{\pi^{0}}\over E_{\pi^{-}}+E_{\pi^{0}}}.
\end{equation}
All pion energies could be taken in laboratory frame. The reason for $y_{1,2}$
choice were the $\tau$ decay  matrix element
properties.

%%%%%%%%%%%%%%%%%%%%%%%%%%%%%%%%%%

\subsection{Toward other decay modes}

\begin{figure}

\begin{minipage}{0.45\linewidth}
%  \centerline{\includegraphics[width=7.5cm,angle=0]{ryski_mixing_025_7}}
   \epsfig{file=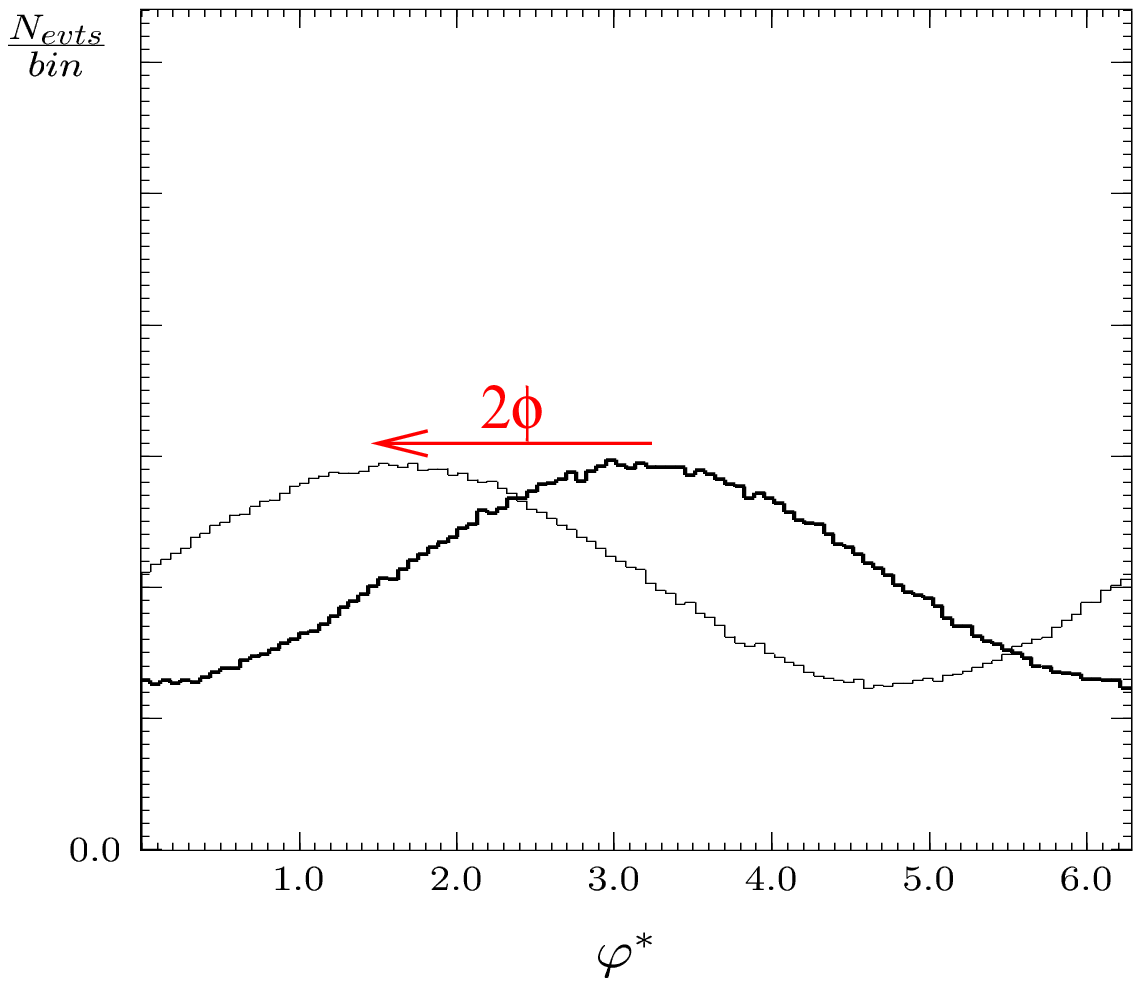,width=70mm,height=60mm}
\end{minipage}
\begin{minipage}{0.45\linewidth}
  %     \centerline{\includegraphics[width=7.5cm,angle=0]{ryski_mixing_025_8}}
     \epsfig{file=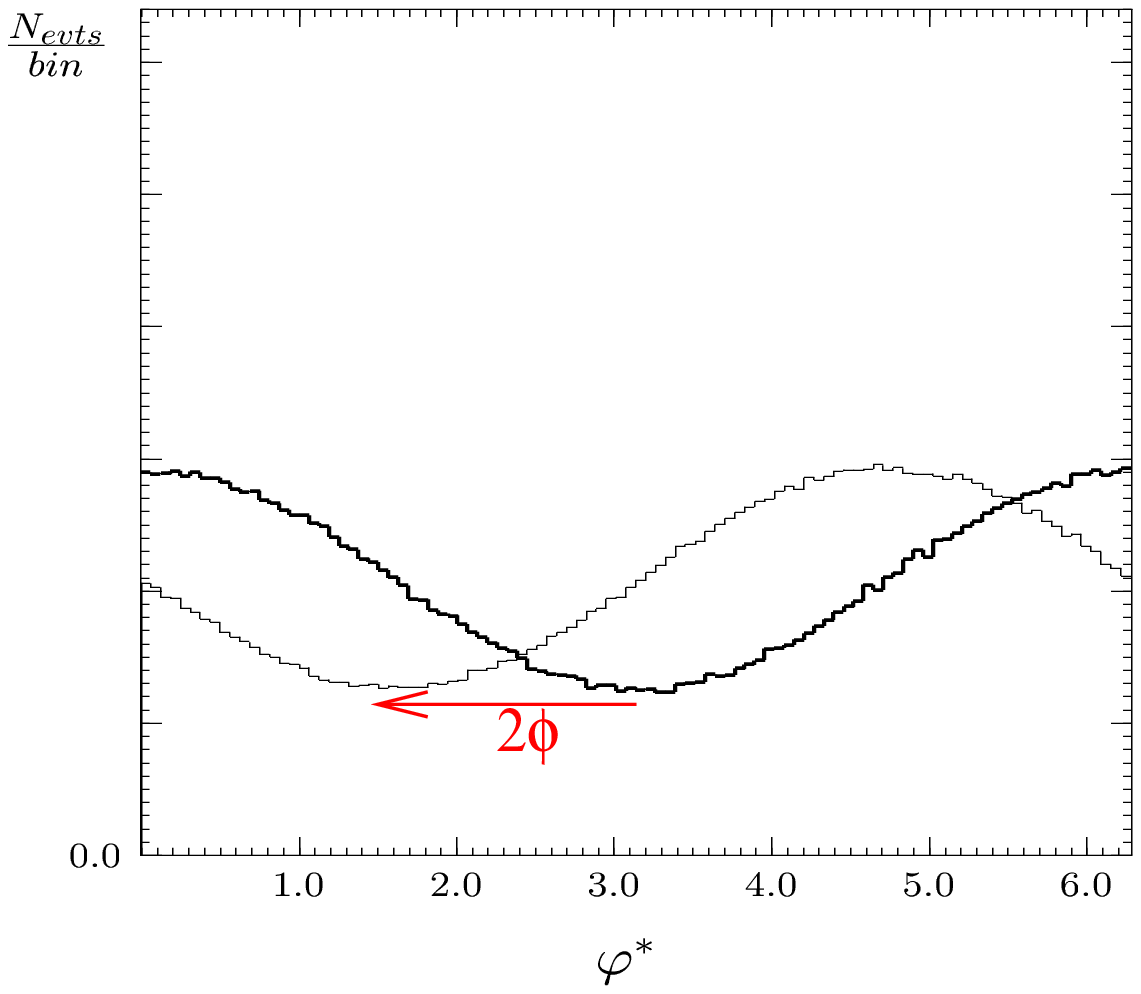,width=70mm,height=60mm}
\end{minipage}
\caption{Distribution of acoplanarity angle $\varphi^*$ for  
  $\rho^+$ and $\rho^-$ decay products planes,
  defined in $\rho^+\rho^-$ pair rest frame, selection on the sign of $y_1 \cdot y_2$ is used, cases of scalar and mixed scalar-pseudoscalar cases are compared.
  \label{fig:acodistr}}
\end{figure}

\begin{figure}
  \begin{center}                               
{
   \includegraphics[width=7.5cm,angle=0]{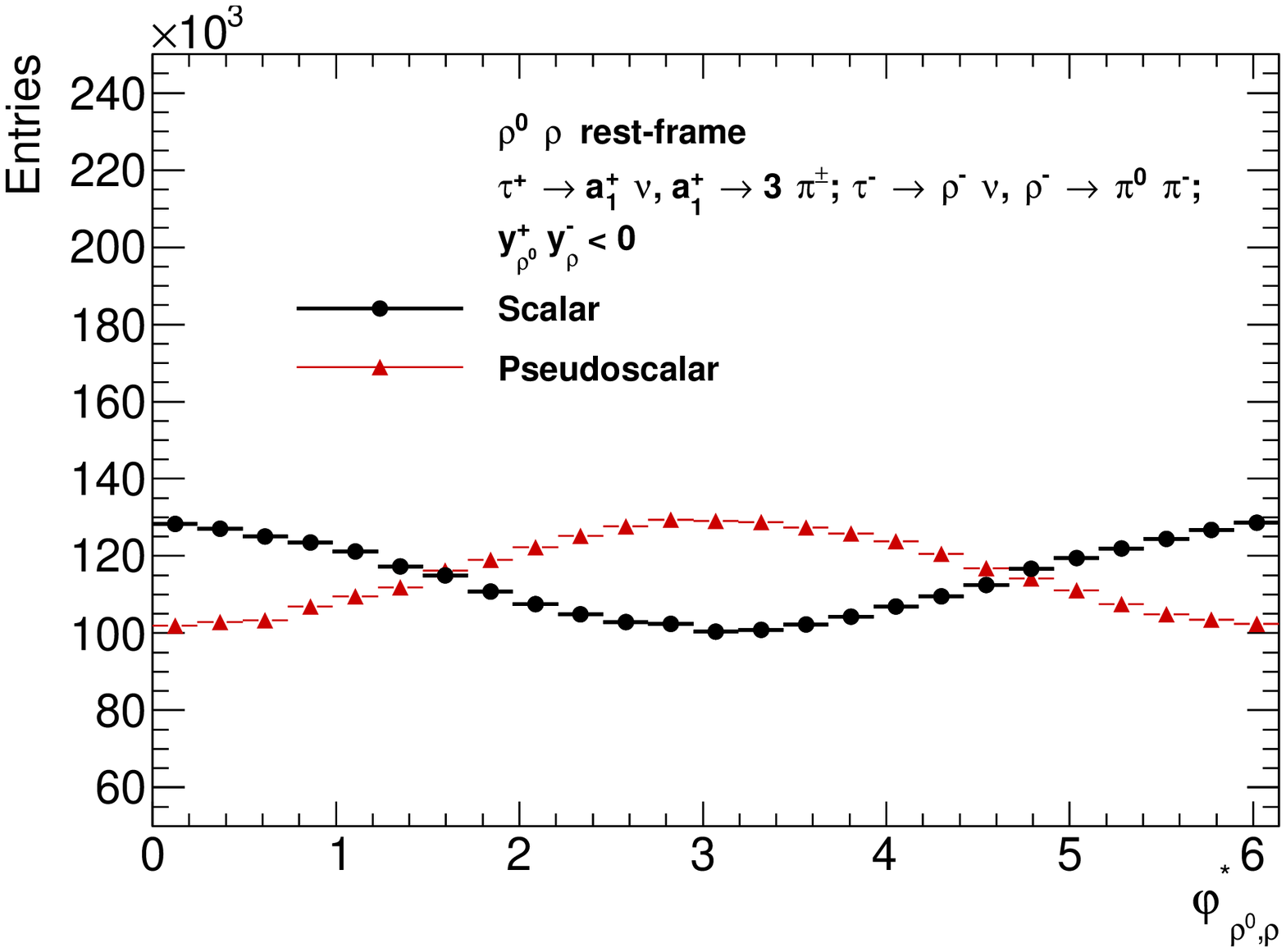}
   \includegraphics[width=7.5cm,angle=0]{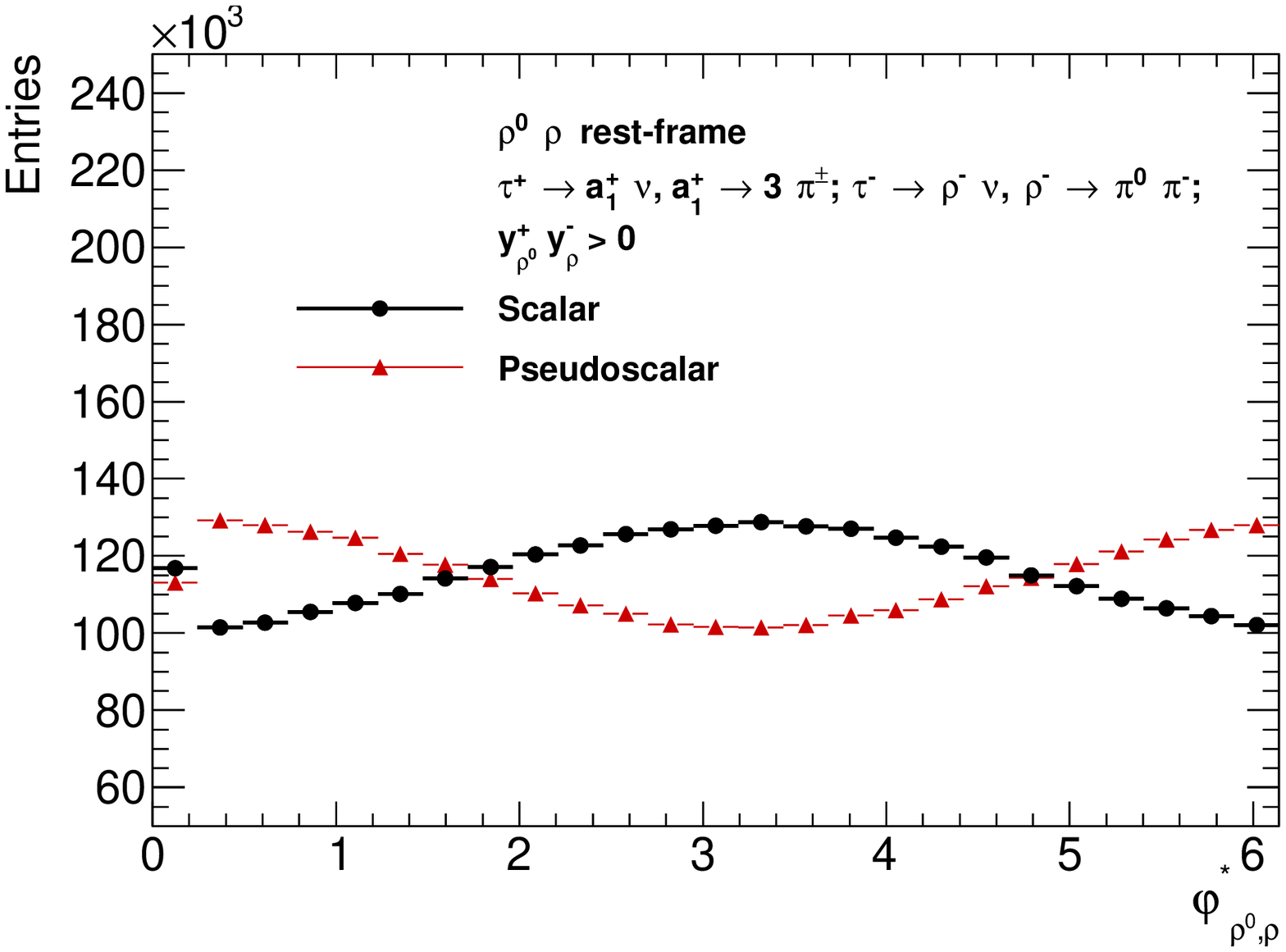}
   \includegraphics[width=7.5cm,angle=0]{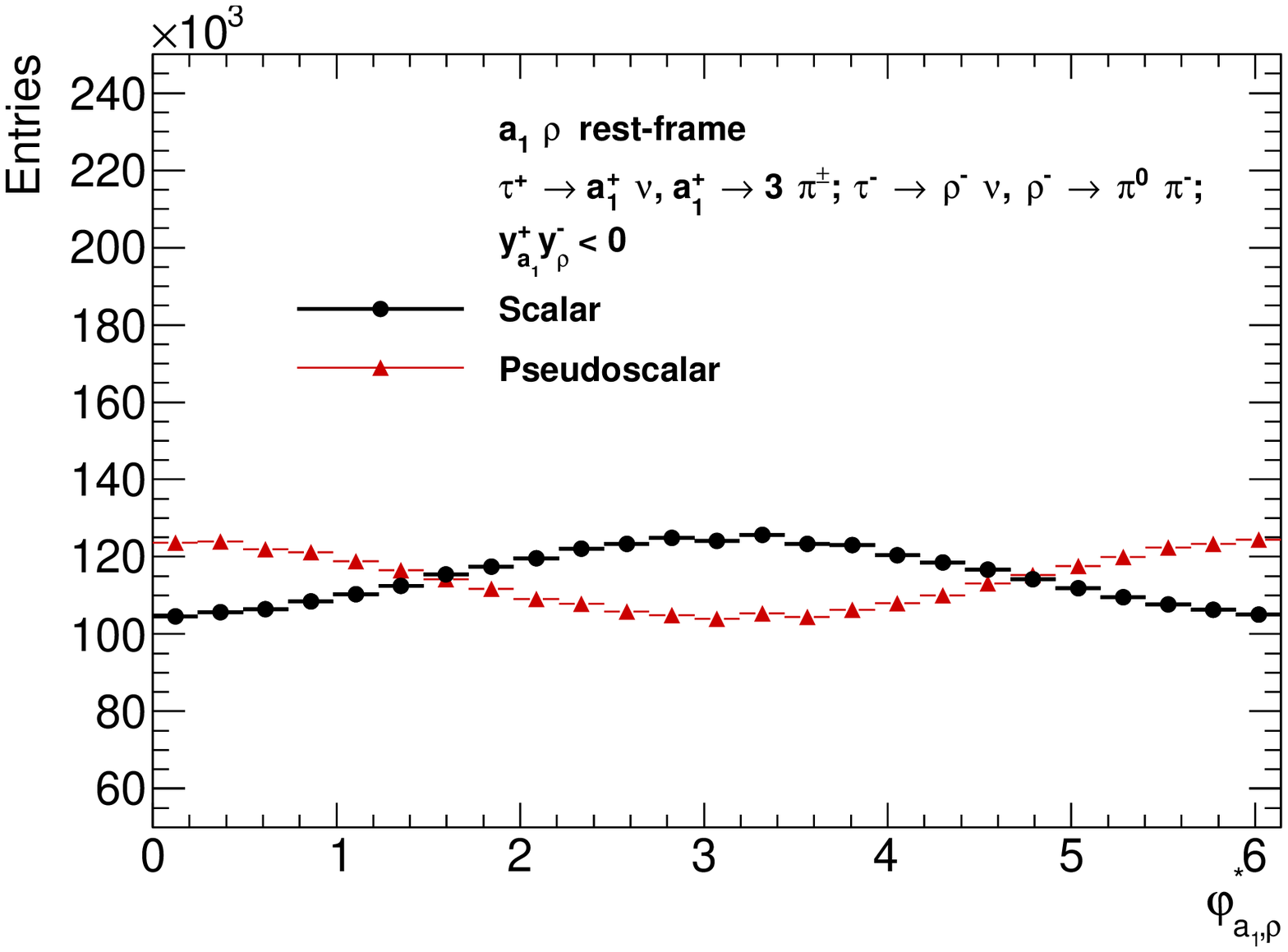}
   \includegraphics[width=7.5cm,angle=0]{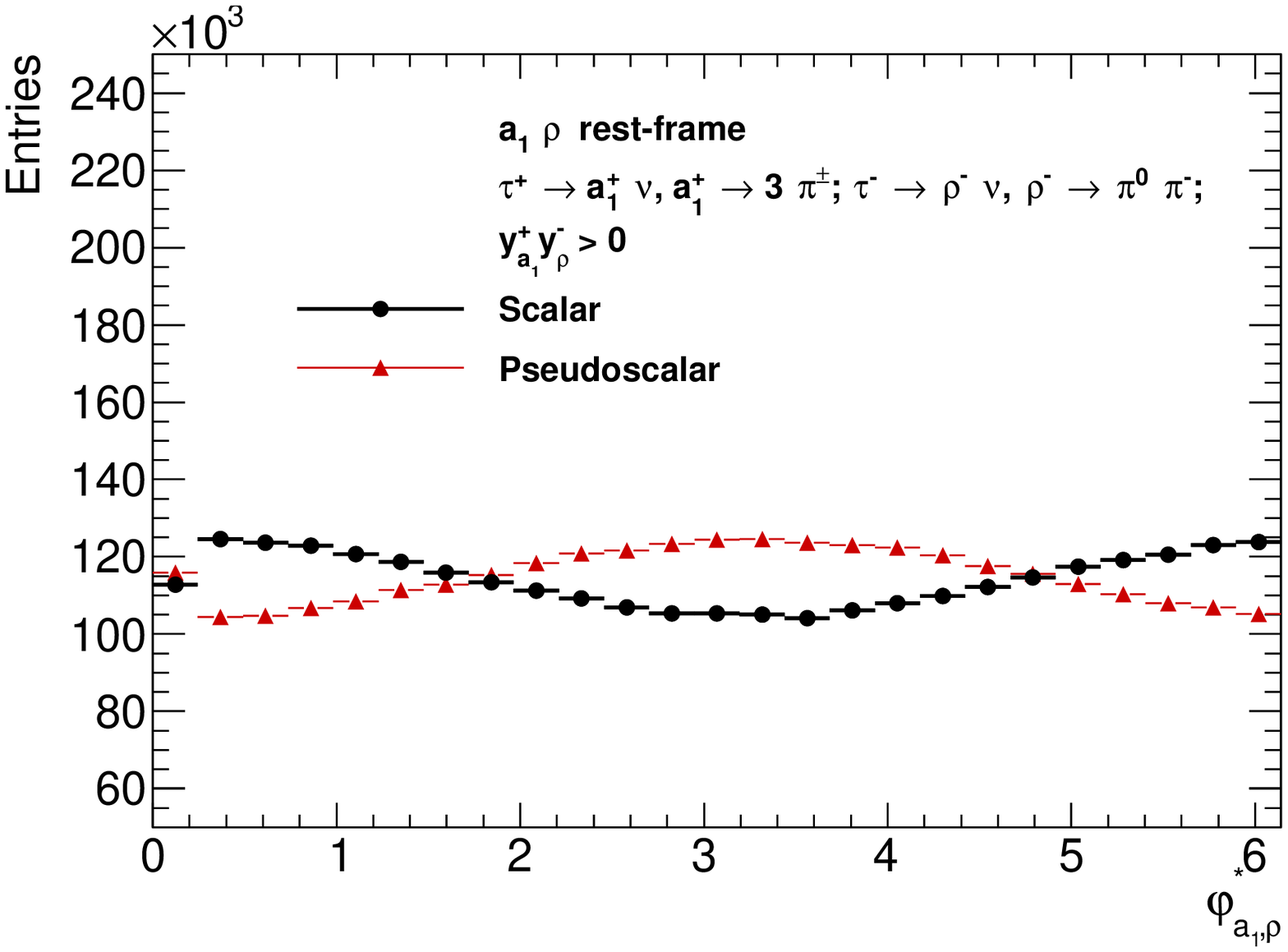}
}
\end{center}
\caption{Acoplanarity angles of oriented half decay planes: $\varphi^*_{\rho^{0} \rho}$ (top), $\varphi^*_{{a_1} \rho}$ (bottom), 
for events grouped by the sign of  $y_{\rho^{0}}^+ y_{\rho}^-$ and $y_{a_1}^{+} y_{\rho}^{-}$ respectively.
\label{fig:a1rho}}
\end{figure}

Even though formally distributions of Fig.~\ref{fig:acodistr} are 
one dimensional
they require  selection with  sign of the $y_1 \cdot y_2$ product.
In fact, already in the presented above case, minor improvement can be achieved if
 the statistical
analysis of the 3-dimensional distribution
over acoplanarity supplemented with $y_1$ and $y_2$ is studied.
%%%%%%%%%%%%%%%%%%%%%%%%%%%%%%%%%%
%%%%%%%%%%%%%%%%%%%%%%%%%%%%%%%%%%
%%%%%%%%%%%%%%%%%%%%%%%%%%%%%%%%%%
This necessity to study multidimensional distribution become even more profound
if one turn attention to $\tau$ decays with 3 pions in final state.
Then not only one plane can be span over the visible decay products, but four,
corresponding to intermediate $a_1$ or $\rho$ decays.
Indeed, each of such planes provide some sensitivity to Higgs CP, as it can be seen from Fig.~\ref{fig:a1rho}, taken from Ref.\cite{Jozefowicz:2016kvz}.
For each plane pair analogous to eq.~(\ref{eq:yi}) selections would need to be studied.
Each such distribution is of rather minor sensitivity to CP and
also because of correlations, such an approach become challenging from
the point of view of statistical analysis. 

We have used  Machine Learning (ML) techniques available as explained
in\cite{deeplearningbook}.
Results of Table \ref{Tab:PhysRevD} taken from Ref.\cite{Jozefowicz:2016kvz}
are encouraging,
they were stable with respect to inclusion of some approximated detector
smearing\cite{Barberio:2017ngd} see Table \ref{table:baseline}.
Statistical uncertainties were derived from a bootstrap method. Systematic uncertainty was calculated with the method also described in\cite{Barberio:2017ngd}.
We played later, in Ref.\cite{Lasocha:2018jcb} with several options of inputs, where so called expert variables, like our acoplanarity angles were used or not.
Depending on the ML variants, the performance varied.
Sometimes sensitivity was completely absent. In general,
structures present in the data, which were of polynomial nature were easy
to recognize, but the one related to boosts, especially strongly relativistic
boosts, especially when rotatory symmetries had to be identified where
challenging for the ML algorithms.

\begin{table}
 \vspace{2mm}
  \begin{center}
    \begin{tabular}{|l|r|r|r|}
    \hline
    Line content  & Channel: $\rho^\pm-\rho^\mp$          &  Channel: $a_1^\pm-\rho^\mp$                      &   Channel: $a_1^\pm-a_1^\mp$ \\
                  &             $\rho^\pm \to \pi^\pm \pi^0$ &  $a_1^\pm \to \rho^0 \pi^\pm, \rho^0 \to \pi^+ \pi^-$ & $a_1^\pm \to \rho^0 \pi^\pm, \rho^0 \to \pi^+ \pi^-$\\
                  &                                         &                     $\rho^0 \to \pi^+ \pi^-$        &               \\
    \hline
    Fraction of $H\to \tau \tau$  &  6.5\% & 4.6\% &   0.8\% \\
    \hline
    Number of features        &  24   & 32    &   48 \\
    \hline
    True (oracle) classification   & 0.782 & 0.782 & 0.782 \\
    ML classification          & 0.638 & 0.590 & 0.557 \\
    \hline
    \end{tabular}
  \end{center}
    \caption{
    The ML performance  for discrimination
    between scalar ad pseudoscalar Higgs CP state. The 4-momenta of hadronic decay products are used only.\label{Tab:PhysRevD} }
 \vspace{2mm}
\end{table}	

\begin{table}
    \begin{tabular}{|c|c|c|c|c|c|c|c|}
    \hline
    \multicolumn{4}{|c|}{Features}  & \multicolumn{1}{c|}{{}{ }{Exact $\pm$ (stat)}} & {}{}{Smeared $\pm$ (stat) $\pm$ (syst)} & \multicolumn{1}{c|}{{}{ }{From Ref.\cite{Jozefowicz:2016kvz}}} \\ \cline{1-4}
    $\phi^*$ & 4-vec &  $y_i$ & $m_i$ & \multicolumn{1}{c|}{}    & \multicolumn{1}{c|}{}             &                      \\
    \hline
    \multicolumn{7}{|c|}{$a_1-\rho$ Decays}\\
    \hline
    \cmark & \cmark & \cmark & \cmark & $0.6035 \pm 0.0005$ & $0.5923 \pm 0.0005 \pm 0.0002$ & 0.596 \\
    \cmark & \cmark & \cmark & -      & $0.5965 \pm 0.0005$ & $0.5889 \pm 0.0005 \pm 0.0002$ & - \\
    \cmark & \cmark & -      & \cmark & $0.6037 \pm 0.0005$ & $0.5933 \pm 0.0005 \pm 0.0003$ & - \\
    -      & \cmark & -      & -      & $0.5971 \pm 0.0005$ & $0.5892 \pm 0.0005 \pm 0.0002$ & 0.590 \\
    \cmark & \cmark & -      & -      & $0.5971 \pm 0.0005$ & $0.5893 \pm 0.0005 \pm 0.0002$ & 0.594 \\
    \cmark & -      & \cmark & \cmark & $0.5927 \pm 0.0005$ & $0.5847 \pm 0.0005 \pm 0.0002$ & 0.578 \\
    \cmark & -      & \cmark & -      & $0.5819 \pm 0.0005$ & $0.5746 \pm 0.0005 \pm 0.0002$ & 0.569 \\
    \hline
    \multicolumn{7}{|c|}{$a_1-a_1$ Decays}\\
    \hline
    \cmark & \cmark & \cmark & \cmark & $0.5669 \pm 0.0004$ & $0.5657 \pm 0.0004 \pm 0.0001$ & 0.573 \\
    \cmark & \cmark & \cmark & -      & $0.5596 \pm 0.0004$ & $0.5599 \pm 0.0004 \pm 0.0001$ & - \\
    \cmark & \cmark & -      & \cmark & $0.5677 \pm 0.0004$ & $0.5661 \pm 0.0004 \pm 0.0001$ & - \\
    -      & \cmark & -      & -      & $0.5654 \pm 0.0004$ & $0.5641 \pm 0.0004 \pm 0.0001$ & 0.553 \\
    \cmark & \cmark & -      & -      & $0.5623 \pm 0.0004$ & $0.5615 \pm 0.0004 \pm 0.0001$ & 0.573 \\
    \cmark & -      & \cmark & \cmark & $0.5469 \pm 0.0004$ & $0.5466 \pm 0.0004 \pm 0.0001$ & 0.548 \\
    \cmark & -      & \cmark & -      & $0.5369 \pm 0.0004$ & $0.5374 \pm 0.0004 \pm 0.0001$ & 0.536 \\
    \hline
    \end{tabular}
    \caption{Area Under Curve  for Neural Network (NN) trained to separate scalar and pseudoscalar hypotheses with combinations of input features marked with a \cmark. 
    Results in the column labeled ``Exact" are from NNs trained with exact data. 
    The results in column labeled ``Smeared" are from NNs trained with smeared data. 
    \label{table:baseline}
    }
\end{table}

\section{Conclusions}
In our studies we could observe that ML solutions were helpful for
significance evaluation in case of Higgs CP signatures in $H \to \tau\tau$
channel. Massively multi-dimensional signatures could have been controlled.
We have identified that features which are related to multi-scale
nature of resonance decays of masses ranging from Higgs of 125 GeV to
$\rho$ meson of 0.7 GeV wre a challenge for ML.  General purpose algorithms such as of
Refs.\cite{Chen:2016btl,Breiman:2001hzm,Bevan:2017stx}
required such adjustment.
It was enough to boost and
rotate input four momenta to appropriate frames, then use of expert variables
was not necessary.
On the other hand, solutions, like studied and developed in\cite{Erdmann:2018shi} which are Lorentz group structure savvy
may not need such pre-conditioning. ML techniques were useful for phenomenology of Higgs CP in
$\tau\tau$ decay channel. For the reversed perspective,  the signatures
offered good investigation ground for ML algorithms. The case was suitable
for such studies, because
Matrix Elements  are available for event weight calculation. 
Analytic dependence  of  weights for  Higgs CP dependent part is
clear. Studies may be thus of broader than just
Higgs CP interest.

On the other hand, even though ML solutions may seem as panacea for all
optimization, one should not ignore Optimal Variables approach, which is
not only useful to indicate which experimental features are
most important for future sensitivity improvements but also provide
essential benchmarks. For example, if ML improvements are surprisingly
promising, this may indicate that some technical imperfections
of Monte Carlo simulations differ e.g. between simulations for signal and background, thus contributing
inappropriately to classification.  

\section*{Acknowledgments}
This project was supported in part from funds of Polish National Science
Centre under decision DEC-2017/27/B/ST2/01391.

\end{document}